\documentclass[12pt]{article}
\usepackage{graphicx}
\newlength{\vshift}
\input epsf.tex
\newlength{\hshift}

\usepackage{amssymb}
\usepackage{amsmath}
\usepackage{amsthm}
\usepackage{amsopn}
\def\uno{\mbox{1 \kern-.59em {\rm l}}}

\def\beq{\begin{equation}}
\def\eeq{\end{equation}}
\def\bea{\begin{eqnarray}}
\def\eea{\end{eqnarray}}

\begin{document}

 \vspace*{3cm}

\begin{center}

{\bf{\Large  Thermodynamic Geometry of Deformed Bosons and Fermions }}

\vskip 4em

{ {\bf Behrouz ~Mirza} \footnote{e-mail: b.mirza@cc.iut.ac.ir}\: and
\: {\bf Hossein ~Mohammadzadeh}
\footnote{e-mail:h.mohammadzadeh@ph.iut.ac.ir}}

\vskip 1em

{$^{1}$ Department of Physics, Isfahan University of Technology, Isfahan,
84156-83111, Iran}\\
$^{2}$ Department of Physics, University of Mohaghegh Ardabili , P.O. Box 179, Ardabil, Iran
 \end{center}

 \vspace*{1.9cm}

\begin{abstract}We construct the thermodynamic geometry of an ideal $q$-deformed boson and fermion gas. We investigate some thermodynamic properties such as the stability and statistical interaction. It will be shown that the statistical interaction of $q$-deformed boson gas is attractive, while it is repulsive for the $q$-deformed fermion one. Also, we will consider the singular point of  the thermodynamic curvature to obtain some new results about the condensation of $q$-deformed bosons and show that there exist a finite critical phase transition temperature even in low dimensions.  It is shown that the thermodynamic curvature of $q$-deformed boson and fermion quantum gases diverges as a power-law function with respect to temperature at zero temperature limit.
\end{abstract}

PACS number(s): 05.20.-y, 05.30.-d, 05.70.Fh
\newpage
\section{Introduction}
The spin-statistics theorem has a fundamental role in connecting the quantum mechanics of many-body systems and quantum statistical mechanics \cite{Pauli,weinberg}. The wave function of a system of identical integer-spin particles  (bosons) is symmetric under exchange, while it is antisymmetric for a system of identical half-integer particles (fermions). The symmetrization and antisymmetrization of the many-body wave function is directly related to the commutation and anticommutation relation of creation and annihilation operators in the language of second quantization. Also, the number of particles occupying the same quantum state and the statistics are mutually connected. However, there is some generalization for the ordinary statistics which is based on the generalization of symmetrization of the many-body system wave function \cite{Leinas,wilczek}, the Pauli exclusion principle \cite{Haldane,polychronakos}, the algebra of commutation and anticommutation relation \cite{Green,Biedenharn,Mcfarlan,Greenberg1,Greenberg2}, and the number of particles occupying the same quantum state \cite{Gentile1,Gentile2}. There is yet, another kind which is based on the generalization of the Boltzmann-Gibbs entropy \cite{Tsallis}.

The thermodynamics of different models can be considered by a qualitative tool, namely the thermodynamic curvature, which has been introduced by the  theory of thermodynamic geometry \cite{Gibbs,Weinhold,Ruppeiner1,Ruppeiner01,brody2,Nulton,Mrugala2,janyszek,brody,Janke,Mrugala3,mirza,Ruppeiner,Ruppeiner2010}.
 Thermodynamic geometry can be used as a measure of statistical interaction, {\it{e.g.}} the thermodynamic curvature of an ideal classical gas is zero and it is  positive and negative for boson and fermion gases, respectively, indicating that the statistical interaction of these models are non interacting, attractive and repulsive \cite{Nulton,Mrugala2,janyszek}. It has been argued that the scalar curvature could be used to show that fermion gases are more stable than boson gases and that, therefore, one can utilize the scalar curvature  as a stability criterion. We notice that the association of the scalar curvature with the stability of the system is somewhat tentative and conjectural and proposed by Janyszek and Mruga{\l}a \cite{Mrugala2,janyszek,Mrugala3}.  Also, it has been shown that the singular point of the thermodynamic curvature coincides with the critical phase transition point of some thermodynamic systems \cite{brody2,janyszek,Janke,Ruppeiner}. Recently, we worked out the thermodynamic curvature of a thermodynamic system in which particles obey fractional exclusion statistics \cite{Mirza1,Mirza2,Mirza3}. The study yielded some interesting results about the thermodynamic properties of these systems.
In the present paper, we mainly investigate the thermodynamic geometry of another kind of intermediate statistic, i.e., $q$-deformed bosons and fermions.
One interesting realization of the intermediate statistic is related to the deformation of  commutation and anti-commutation relations, which has led to the $q$-deformed algebra of creation and annihilation operators and is  related to the general theory of quantum groups \cite{Biedenharn,Mcfarlan}. Many authors have considered the statistics and thermodynamics of such models \cite{Lee,Ng,chaichian,Su,Rubin,Swamy,Lavagno1,Ubriaco,Lavagno2,Lavagno3,Lavagno4}. We construct the thermodynamic geometry of $q$-deformed boson and fermion gases by using the $q$-calculus  introduced by Jackson \cite{Jackson}.

The rest of paper is organized as follows. In Sec. II, we summarize the $q$-calculus and deformed algebra. We derive the internal energy and particle number of $q$-deformed bosons and fermions based on the Jackson derivative (JD) of the logarithm of partition function. In Sec. III, we construct the thermodynamic geometry of $q$-deformed bosons and fermions. Finally, we consider the singular point of thermodynamic geometry to show that the singular point coincides with the critical point, where a phase transition such as Bose-Einstein condensation occurs.

\section{Deformed Algebra and $q$-calculus }
The symmetric $q$-oscillators algebra is defined by the following equations \cite{Lee,Ng,chaichian},
 \bea
 cc^{\dag}-kqc^{\dag}c=q^{-N}
 \eea
 \bea\label{algebra}
 &&{[c,c]_{k}}=[c^{\dag},c^{\dag}]_{k}=0,\nonumber\\
 &&{[N,c^{\dag}]}=c^{\dag},~~~~~~~~[N,c]=-c,\\
 &&c^{\dag}c=[N],~~~~~~~~~~~cc^{\dag}=[1+k N],\nonumber
 \eea
where, $[A,B]_{x}=AB-xBA$, and $k=1$ stands for symmetric $q$-bosons while $k=-1$ for symmetric $q$-fermions. Also, $c,c^{\dag}$ and $N$ are the creation, annihilation, and number operator, respectively. In the generalized $q$-deformation, the $q$-basic number and the Jackson derivative of real functions are defined as
 \bea
 &&[x]=\frac{q^{x}-q^{-x}}{q-q^{-1}},\\
 &&{\cal{D}}^{(q)}_{x}f(x)=\frac{f(q x)-f(q^{-1}x)}{(q-q^{-1})x}.
  \eea
Now, consider the following Hamiltonian of non-interacting $q$-deformed bosons and fermions
 \bea
 H=\sum_{i}(\epsilon_{i}-\mu)N_{i},
 \eea
where, $\epsilon_{i}$ is the kinetic energy in each state, $\mu$ is the chemical potential and $N_i$ is the number operator, which is deformed according to Eqs. (\ref{algebra}).  Mean occupation number is defined by
 \bea
 [n_i]=\frac{\mathrm{Tr}(e^{-\beta H}c^{\dag}_{i}c_{i})}{{\mathrm{Tr}(e^{-\beta H})}}
 \eea
Applying the cyclic property of the trace and using the $q$-deformed algebra, one can obtain
 \bea
 [n_{i}]=e^{-\eta}[kn_{i}+1],
 \eea
 where, $\eta=\beta(\epsilon_{i}-\mu)$. Using algebraic calculations, one can evaluate the mean occupation number as \cite{Biedenharn,Mcfarlan,Rubin}
 \bea
 {n}_{i}=\frac{1}{q-q^{-1}}\ln\left(\frac{z^{-1}e^{\beta\epsilon_{i}}-kq^{-k}}{z^{-1}e^{\beta\epsilon_{i}}-kq^{k}}\right).\label{NG}
 \eea
Now, we can start with the logarithm of the grand partition function  of the $q$-deformed boson and fermion, which is given by
 \bea
   \ln{\cal{Z}}=-k\sum_{i}\ln(1-kze^{-\beta\epsilon_{i}}).
 \eea
We note that the Hamiltonian is designated to be a linear function of the number operator. It is straightforward to show that the thermodynamic quantities in ordinary statistics such as boson or fermion gas are given by
 \bea
 U&=&-(\frac{\partial}{\partial\beta}\ln{\cal{Z}})_{\gamma}=\epsilon_{i}(\frac{\partial}{\partial y_{i}}\ln{\cal{Z}})_{z}=\sum_{i}\epsilon_{i}n_{i},\nonumber\\
 N&=&-(\frac{\partial}{\partial\gamma}\ln{\cal{Z}})_{\beta}=z(\frac{\partial}{\partial z}\ln{\cal{Z}})_{y_{i}}=\sum_{i}n_{i},
 \eea
where, $z=e^{\mu/kT}=e^{-\gamma}$ is the fugacity and $y_{i}=e^{-\beta\epsilon_{i}}$. It has been shown that by replacing the ordinary derivative with JD, one can find a simple form for the number of particles in the $q$-deformed case \cite{Lavagno2,Lavagno3,Lavagno4}. In other words, we define the JD derivatives with respect to the thermodynamic parameters ($\beta,\gamma$) by
 \bea
 {\cal{D}}^{(q)}_{\beta}=-\epsilon_{i}{\cal{D}}^{(q)}_{y_{i}},~~~~~{\cal{D}}^{(q)}_{\gamma}=-z{\cal{D}}^{(q)}_{z},
 \eea
 and immediately the well defined particle number and internal energy is obtained by
 \bea
U&=&-{\cal{D}}^{(q)}_{\beta}\ln{\cal{Z}}=-k\sum_{i}\epsilon_{i}{\cal{D}}^{(q)}_{y_{i}}\ln(1-kze^{-\beta\epsilon_{i}})\mid_{_{z}}=\sum_{i}\epsilon_{i}n_{i}, \nonumber\\
 N&=&-{\cal{D}}^{(q)}_{\gamma}\ln{\cal{Z}}=z{\cal{D}}^{(q)}_{z}\ln{\cal{Z}}\mid_{_{y_{i}}}=\sum_{i}n_{i},
 \eea
where, $n_{i}$ has been expressed in Eq. (\ref{NG}). In the thermodynamic limit and in a $D$ dimensional box of volume $L^D$ with particles satisfying the  dispersion relation $ \epsilon=ap^\sigma$ where $\sigma$ takes constant values for example $\sigma=2$ for non relativistic and $\sigma=1$ for ultra relativistic particles,the summation can be replaced by the integral.
 \bea
 \sum_{i}\longrightarrow\frac{A^D}{\Gamma(\frac{D}{2})}\int_{0}^{\infty}\epsilon^{D/\sigma -1}d\epsilon,\label{SI}
 \eea
where, $A=(2/\sigma)^{1/D}\frac{L\sqrt{\pi}}{a^{1/\sigma} h}$ is a constant which for
simplicity, we will set it practically equal to one $(A=1)$ later. For the symmetric $q$-deformed statistics, we can work out the thermodynamic quantities using the mean occupation number in Eq. (\ref{NG}),
  \bea
 \label{GQ}{U}&=&\frac{A^D}{q-q^{-1}}\frac{1}{{\Gamma(\frac{D}{2})}}\int_{0}^{\infty}\epsilon^{\frac{D}{\sigma}}\ln\left(\frac{z^{-1}e^{\beta\epsilon}-kq^{-k}}{z^{-1}e^{\beta\epsilon}-kq^{k}}\right)d\epsilon\nonumber\\
 &=&A^D\frac{\Gamma(\frac{D}{\sigma}+1)}{\Gamma(\frac{D}{2})}\frac{\beta^{-(\frac{D}{\sigma}+1)}}{q-q^{-1}} {H}_{\frac{D}{\sigma}+2}(z,k,q)\nonumber\\
 {N}&=&\frac{A^D}{q-q^{-1}}\frac{1}{{\Gamma(\frac{D}{2})}}\int_{0}^{\infty}\epsilon^{\frac{D}{\sigma}-1}\ln\left(\frac{z^{-1}e^{\beta\epsilon}-kq^{-k}}{z^{-1}e^{\beta\epsilon}-kq^{k}}\right)d\epsilon\nonumber\\
 &=&A^D\frac{\Gamma(\frac{D}{\sigma})}{\Gamma(\frac{D}{2})}\frac{\beta^{-(\frac{D}{\sigma})}}{q-q^{-1}} {H}_{\frac{D}{\sigma}+1}(z,k,q),
 \eea
where,
 \bea
 {H}_{n}(az,k,q)=Li_{n}(azkq^{k})-Li_{n}(azkq^{-k}),
 \eea
 where, $Li_{n}(x)$ denotes the polylogarithm function.

\section{Thermodynamic curvature of a deformed statistic gas}
The geometrical study of the phase space of statistical
thermodynamics dates back to J. W. Gibbs \cite{Gibbs} . The geometrical
thermodynamics was developed by Weinhold and Ruppeiner
\cite{ Weinhold,Ruppeiner1}. Ruppeiner geometry is based on the entropy representation, where
we denote the extended set of $n+1$ extensive variables of the
system by $X=(U,N^{1},...,V,...,N^{r})$, while
Weinhold worked in the energy representation in which the
extended set of $n+1$ extensive variables of system are denoted
by $Y=(S,N^{1},...,V,...,N^{r})$ \cite{Ruppeiner01}. It should be
noted that we can work in any thermodynamic potential
representation that is the Legendre transform of the entropy or
the internal energy. The metric of this representation may be the
second derivative of the thermodynamic potential with respect
 to intensive variables; for example, the thermodynamic potential
$\Phi$ which is defined as,
    \bea
    \Phi=\Phi(\{F^{i}\})
    \eea
where, $F=(1/T,-\mu^{1}/T,...,P/T,...,-\mu^{r}/T)$. $\Phi$ is the
Legendre transform of entropy with respect to the extensive
parameter $X^{i}$,
    \bea
    F^{i}=\frac{\partial S}{\partial X^{i}}.
    \eea
The metric in this representation is given by
    \bea
    G_{ij}=\frac{\partial^{2}\Phi}{\partial F^{i}\partial F^{j}}.
    \eea
Janyszek and Mruga{\l}a used the partition function to introduce
the metric geometry of the parameter space \cite{Mrugala2},
    \bea\label{M1}
    G_{ij}=\frac{\partial^{2}\ln\cal{Z}}{\partial \beta^{i}\partial\beta^{j}}
    \eea
where, $\beta^{i}=F^{i}/k$ and ${\cal{Z}}$ is the partition function.

According to Eqs. (\ref{GQ}), the parameter space of an ideal
$q$-deformed gas is  $(\beta,\gamma)$, where $
{\beta} = {1 / kT}$ and $\gamma = -\mu/ kT $ . For computing the
thermodynamic metric, we select one of the extended variables as
the constant system scale. We will implicitly pick volume in
working with the grand canonical ensemble \cite{Mrugala2}. We can
evaluate the metric elements by the metric definition in Eq. ({\ref{M1}). But, there is a subtle point to notice. According to the definition of thermodynamic quantities such as the particle number and internal energy, we have to replace the ordinary derivative by JD to obtain the correct $q$-deformed quantities. Therefore, naturally we define the metric element by the second JD of the logarithm of the partition function. Also, the first JD of the logarithm of the partition function yields internal energy and particle number. Thus, the metric elements of the
thermodynamic space of an ideal $q$-deformed gas are given
by
   \bea
     \label{vq}
     G_{\beta\beta}&=&{\cal{D}}^{(q)}_{\beta}({\cal{D}}^{(q)}_{\beta}\ln{\cal{Z}})=
     -{\cal{D}}^{(q)}_{\beta}U=\sum_{i}\epsilon^{2}_{i}{\cal{D}}^{(q)}_{y_{i}}n_{i}\nonumber\nonumber\\
     G_{\beta\gamma}&=&{\cal{D}}^{(q)}_{\beta}({\cal{D}}^{(q)}_{\gamma}\ln{\cal{Z}})=
     -{\cal{D}}^{(q)}_{\beta}N=\sum_{i}\epsilon_{i}{\cal{D}}^{(q)}_{y_{i}}n_{i}\\
     G_{\gamma\gamma}&=&{\cal{D}}^{(q)}_{\gamma}({\cal{D}}^{(q)}_{\gamma}\ln{\cal{Z}})=-{\cal{D}}^{(q)}_{\gamma}N=z\sum_{i}{\cal{D}}^{(q)}_{z}n_{i}.\nonumber
   \eea
Replacing the summation by the integral according to Eq. (\ref{SI}) and evaluating the JD of mean occupation number in Eq. (\ref{NG}), we will have the following relations:
\bea
 G_{\beta\beta}&=&\frac{\Gamma(\frac{D}{\sigma}+2)}{\Gamma(\frac{D}{2})}\frac{\beta^{-(\frac{D}{\sigma}+2)}}{(q-q^{-1})^{2}}
 \left(H_{{\frac{D}{\sigma}+3}}(z q,k,q)-H_{{\frac{D}{\sigma}+3}}(z q^{-1},k,q)\right),\nonumber\\
 G_{\beta\gamma}&=&\frac{\Gamma(\frac{D}{\sigma}+1)}{\Gamma(\frac{D}{2})}\frac{\beta^{-(\frac{D}{\sigma}+1)}}{(q-q^{-1})^{2}}
 \left(H_{{\frac{D}{\sigma}+2}}(z q,k,q)-H_{{\frac{D}{\sigma}+2}}(z q^{-1},k,q)\right),\\
  G_{\gamma\gamma}&=&\frac{\Gamma(\frac{D}{\sigma})}{\Gamma(\frac{D}{2})}\frac{\beta^{-(\frac{D}{\sigma})}}{(q-q^{-1})^{2}}
 \left(H_{{\frac{D}{\sigma}+1}}(z q,k,q)-H_{{\frac{D}{\sigma}+1}}(z q^{-1},k,q)\right).\nonumber
 \eea

We have evaluated the metric elements of the two-dimensional thermodynamic geometry of a $q$-deformed gas. For evaluating the Ricci scalar ($R$) of this geometry, one can use the well-known definition of the Riemann ($R_{ijk}^{l}$) and Ricci ($R_{ij}$) tensors,
 \bea\label{E1}
 && R_{ijk}^{l}=\partial_{j}\Gamma_{k i}^{l}
 -\partial_{k}\Gamma_{j i}^{l}+\Gamma_{k i}^{m}\Gamma_{jm}^{l}
 -\Gamma_{ji}^{m}\Gamma_{km}^{l},\nonumber\\
 &&R_{ij}= R_{ikj}^{k},\\
 && R=G^{ij}R_{ij},\nonumber
  \eea
  where, $\Gamma_{ijk}$ is the Christoffel symbol defined by
 \bea
 \Gamma_{ijk}=\frac{1}{2}\left(G_{ij,k}+G_{ik,j}-G_{jk,i}\right),
 \eea
  where, $G_{ij,k}=\partial_{k}G_{ij}$, and the  two-dimensional  parameter space is defined by  $\beta$ and $\gamma$. It should be noted that we do not use Jackson derivative for constructing a geometrical picture; however, we constructed the metric elements by Jackson derivative of the partition function. The sign convention for $R$ is
arbitrary; so $R$ may be either negative or positive for any
case. Our selected sign convention is the same as that of Janyszek
and Mruga{\l}a, but different from \cite{Ruppeiner01}.
We have used MAPLE for calculating the thermodynamic curvature. The results are provided in the relevant diagrams. Figure \ref{FIG1} shows the thermodynamic curvature of the symmetric three-dimensional non-relativistic ideal $q$-deformed boson gas ($k=1$) as a function of fugacity in isotherm processes for some value of the deformation parameter. It is  observed that the thermodynamic curvature is positive in the full physical range for an arbitrary value of deformation parameter. An interesting phenomenon is observed in $q$-deformed bosons at a specified value of fugacity so that we can argue that there is an upper bound on the  fugacity of $q$-deformed boson gas where  the thermodynamic curvature is singular. Also, it is obvious from Fig. \ref{figure01} that the thermodynamic curvature has the symmetric property ($q\leftrightarrow q^{-1}$) and that, for a specified value of fugacity, we can find two different deformed boson gases with $q<1$ and $q>1$ but with the same value of thermodynamic curvature. We note that the value of thermodynamic curvature for any deformation parameter increases with increasing  fugacity and that specially for the boson gas ($q=1$), this value goes to infinity at the related upper bound of fugacity ($z=1$).
\begin{figure}[t]
    \center
    \includegraphics[width=0.65\columnwidth]{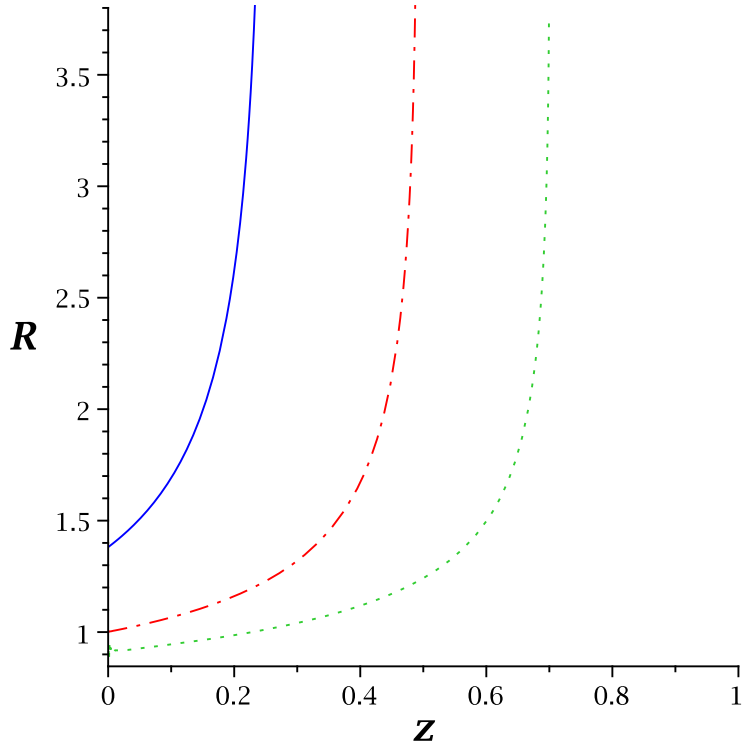}\\
    \caption{ The thermodynamic curvature of a $3D$ symmetric $q$-deformed boson gas with respect to fugacity for an isotherm ($\beta=1$) and some values deformation parameter $(q=0.5\equiv2)$(blue solid line), $(q=0.7\equiv1/0.7)$ (red dash-dot line) and $(q=\sqrt{0.7}\equiv 1/\sqrt{0.7}$ (green dot line).}\label{FIG1}
   \end{figure}
\begin{figure}[t]
    \center
    \includegraphics[width=0.75\columnwidth]{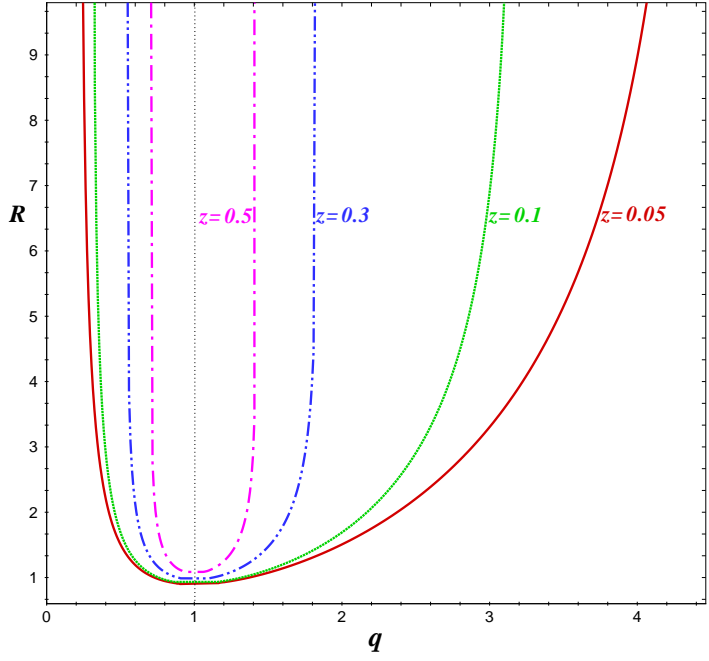}\\
    \caption{The thermodynamic curvature of a $3D$ symmetric $q$-deformed boson gas with respect to deformation parameter for an isotherm ($\beta=1$) and some values of fugacity $(z=0.05)$(red solid line), $(z=0.1)$ (green dotted line), $(z=0.3)$ (blue dash-double dotted line) and $z=0.5$  (purple dash-dotted line). It is obvious that the related deformation parameter at singular point for any value of fugacity is given by $q=\sqrt{z}$ for ($q<1$) and $q=1/\sqrt{z}$ where ($q>1$).}\label{figure01}
   \end{figure}

 Based on the interpretation of the thermodynamic curvature as a measure of statistical interaction, we can argue the $q$-deformed boson gas, as in the case of the boson gas, has an attractive statistical interaction. The stability interpretation of the thermodynamic curvature which is only a conjecture, predicts that the $q$-deformed boson gas may be more stable for larger (smaller) values of the deformation parameter for $q<1$ ($q>1$).

 Another important point concerns the singular point of thermodynamic curvature. We can deduce that for any value of the deformation parameter there is a critical fugacity $z_{q}$ where, $R$ is singular. This critical value of fugacity is given by
   \bea
   z_q=\left\{
         \begin{array}{cc}
           q^{2} & q<1 \\
           q^{-2} & q>1 \\
         \end{array}\right.\label{zq}
    \eea

 Similar calculations have been performed for the standard $q$-deformed boson gas, which is defined in \cite{Lavagno2} and is known in the literature, yielding results quite similar to those for the symmetric one. The only difference between these two $q$-deformed bosons is in their critical fugacity. In the standard case, the following result can be obtained from applying  our
 proposed method or others \cite{Lavagno2},
   \bea
   z_q=\left\{
         \begin{array}{cc}
           1 & q<1 \\
           q^{-2} & q>1 \\
         \end{array}\right.,
    \eea

By setting ($k=-1$), we can investigate the thermodynamic curvature of $q$-deformed fermion gas. Again, we restrict ourself to the three-dimensional non-relativist ideal gas. The results are given in Fig. \ref{FIG2}. The thermodynamic curvature of a $q$-deformed fermion gas is negative for any arbitrary value of deformation parameter in the full physical range, which indicates that the statistical interaction of the $q$-deformed fermion gas is repulsive as is the case with the fermion gas.
\begin{figure}[t]
    \center
    \includegraphics[width=0.65\columnwidth]{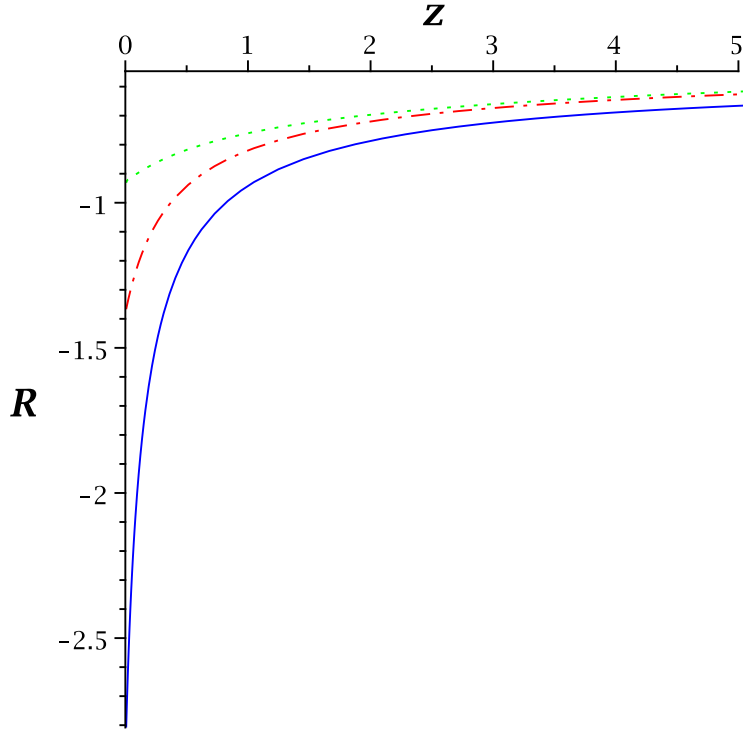}\\
    \caption{The thermodynamic curvature of a $3D$ symmetric $q$-deformed fermion gas with respect to fugacity for an isotherm ($\beta=1$) and some values deformation parameter $(q=0.3\equiv1/0.3)$(blue solid line), $(q=0.5\equiv2)$ (red dash-dot line) and $(q=0.8 \equiv 1.25)$ (green dot line).}\label{FIG2}
   \end{figure}
Also, based on the stability interpretation, we conclude that the $q$-deformed fermion gas may be more stable than the $q$-deformed boson for an arbitrary value of $q$. We have calculated the thermodynamic geometry for  $q$-deformed boson and fermion gases in other dimensions and in the ultra-relativistic limit. Almost similar results have been obtained (not shown here to save space).


We started with Jackson derivatives (JD) to construct the metric elements. If one  insists on  replacing  JD with the ordinary derivative, a new q-deformed geometry may then be developed by using JD to obtain a q-deformed Ricci scalar. Jackson derivative  of the metric elements are given in the following equations for an ideal $q$-deformed boson gas,
    \bea
     G_{\beta\beta,\beta}&=&\frac{\Gamma(\frac{D}{\sigma}+3)}{\Gamma(\frac{D}{2})}\frac{\beta^{-(\frac{D}{\sigma}+3)}}{(q-q^{-1})^{3}}
 \left(2H_{{\frac{D}{\sigma}+4}}(z ,k,q)-H_{{\frac{D}{\sigma}+4}}(z q^{2},k,q)-H_{{\frac{D}{\sigma}+4}}(zq^{-2} ,k,q)\right),\nonumber\\
    G_{\beta\beta,\gamma}&=&\frac{\Gamma(\frac{D}{\sigma}+2)}{\Gamma(\frac{D}{2})}\frac{\beta^{-(\frac{D}{\sigma}+2)}}{(q-q^{-1})^{3}}
 \left(2H_{{\frac{D}{\sigma}+3}}(z ,k,q)-H_{{\frac{D}{\sigma}+3}}(z q^{2},k,q)-H_{{\frac{D}{\sigma}+3}}(zq^{-2} ,k,q)\right),\nonumber\\
   G_{\gamma\gamma,\beta}&=&\frac{\Gamma(\frac{D}{\sigma}+1)}{\Gamma(\frac{D}{2})}\frac{\beta^{-(\frac{D}{\sigma}+1)}}{(q-q^{-1})^{3}}
 \left(2H_{{\frac{D}{\sigma}+2}}(z ,k,q)-H_{{\frac{D}{\sigma}+2}}(z q^{2},k,q)-H_{{\frac{D}{\sigma}+2}}(zq^{-2} ,k,q)\right),\nonumber\\
    G_{\gamma\gamma,\gamma}&=&\frac{\Gamma(\frac{D}{\sigma})}{\Gamma(\frac{D}{2})}\frac{\beta^{-(\frac{D}{\sigma})}}{(q-q^{-1})^{3}}
 \left(2H_{{\frac{D}{\sigma}+1}}(z ,k,q)-H_{{\frac{D}{\sigma}+1}}(z q^{2},k,q)-H_{{\frac{D}{\sigma}+1}}(zq^{-2} ,k,q)\right).\nonumber\\
 \eea
It is straightforward to calculate a q-deformed thermodynamic curvature. This q-deformed Ricci scalar  has  properties similar to the usual scalar curvature
 such as positivity for $q$-deformed bosons and negativity for $q$-deformed fermions. The only difference will be in  the upper bound of fugacity, which will be
   \bea
   z_q=\left\{
         \begin{array}{cc}
           q^{3} & q<1 \\
           q^{-3} & q>1 \\
         \end{array}\right..
    \eea
 Although a $q$-deformed geometry seems to be interesting from theoretical and mathematical viewpoints, no appropriate physical model is available.
\section{Singular point and condensation}
 It has been shown that the singular point of thermodynamic geometry  coincides with the phase transition in several thermodynamic systems. For example, the thermodynamic geometry of Van der Walls gas and that of the Ising model are singular at phase transition points. Also, for an ideal boson gas, the Bose-Einstein condensation occurs at $z=1$, where the thermodynamic curvature is singular. Recently, we have shown that for a kind of fractional exclusion statistics,  the singular point of thermodynamic curvature corresponds to the condensation of a non-pure bosonic system \cite{Mirza3}. Now, we investigate the singular point of the $q$-boson gas.  It has already been shown \cite{Lavagno2,Lavagno3} that the definition of heat capacity leads to  an upper bound on the fugacity of $q$-deformed boson gas, while there is no bound on the fugacity of $q$-deformed fermion gas. We expect a condensation to appear where the upper bound of fugacity for a $q$-deformed boson gas coincides with  the singular point of the thermodynamic curvature (Eq. (\ref{zq})). In this formulation, it is straightforward to obtain the phase transition temperature. We restrict ourselves  to the $q<1$ cases, which is simply extendable to the $q>1$. According to the particle number in Eq. (\ref{GQ}), the critical temperature $T_c$ at the phase transition point is given by,
  \bea
  k_{B}T_{c}=(\frac{\sigma}{2})^{\sigma/D}\frac{ah^\sigma}{\pi^{\sigma/2}}\left[\frac{n\Gamma(D/2)}{\Gamma(D/\sigma)}
  \frac{q-q^{-1}}{H_{\frac{D}{\sigma}+1}(q^{2},1,q)}\right]^{\sigma/D},\label{zzz}
  \eea
 where, $n=N/L^D$ is the density of particles and assumed to be constant.
It has been shown that there is a condensation for the  boson gas at a  finite temperature in the three dimensional space with non-relativistic particles, while the phase transition point goes to zero in the two-dimensional space. Generally, Bose-Einstein condensation does not occur at $D/\sigma\leq 1$ for an ideal boson gas. We investigate the possibility of condensation for an ideal $q$-deformed boson gas in various cases. The important parts of Eq. (\ref{zzz}), which shows the dependency of temperature on dimension, are the  dispersion relation and the  deformation parameter:
 \bea
 A(D,\sigma,q)=\frac{q-q^{-1}}{H_{\frac{D}{\sigma}+1}(q^{2},1,q)}=\frac{q-q^{-1}}{Li_{\frac{D}{\sigma}+1}(q^{3})-Li_{\frac{D}{\sigma}+1}(q)}.
 \eea
 It is interesting that where $q\neq1$, $A(D,\sigma,q)$ has a finite value for any  dimension. In the boson gas limit, this function reduces to
 \bea
 \lim_{q\rightarrow 1}{A(D,\sigma,q)}=\frac{1}{Li_{\frac{D}{\sigma}}(1)},
 \eea
 which is zero for $D/\sigma\leq 1$. In other words, for general values of the deformation parameter, one can find a condensation in all dimensions with non-relativistic or ultra-relativistic particles, while for the boson gas $(q=1)$, the condensation occurs in three (and higher) dimensions and for ultra-relativistic particles, in two dimensions \cite{Chen,Cai,Lavagno5}.
\begin{figure}[t]
    \center
    \includegraphics[width=0.75\columnwidth]{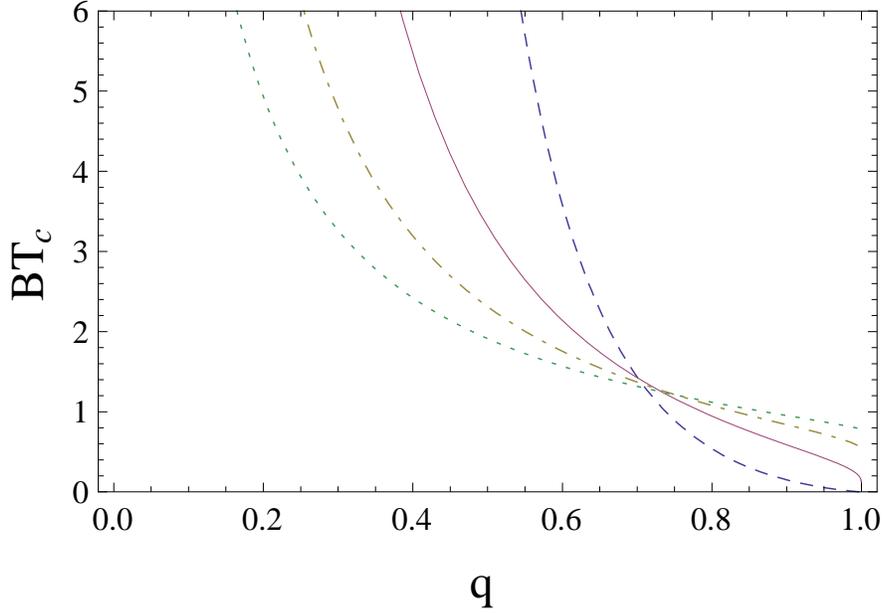}\\
    \caption{Critical temperature as function of deformation parameter in $D=1$ (blue dashed line), $D=2$ (purple solid line),
    $D=3$ (dash-dotted line) and $D=4$ (dotted line) for non relativistic particles. $B=\frac{k_{B}\pi}{n^{2/D}ah^{2}}$ is a constant.}\label{Fig4}
   \end{figure}
Figure (\ref{Fig4}) demonstrates the critical temperature as a function of deformation parameter in various dimensions  for non-relativistic particles. One can find that, in all dimensions, there exists a finite critical temperature for a q-deformed boson gas. The exception is one and two dimensional cases in the limit of boson gas $(q=1)$. Also, it is obvious that the finite critical temperature reduces when the deformation parameter increases and that in some cases, it goes to zero, which excludes condensation in such cases.

According to the general form of the metric elements and their derivatives, the thermodynamic curvature of deformed boson and fermion gases is a function of $\beta$ and $z$ in the following general form
 \bea
 R=\beta^{\frac{D}{\sigma}}f(z,q,D,\sigma).
 \eea
Therefor, in the low temperature limit and at a fixed fugacity, the thermodynamic curvature diverge as $R\simeq T^{-\frac{D}{\sigma}}$. This is a general behavior for quantum gases such as boson, fermion, fractional exclusion statistics, Gentile statistics, as well as $q$-deformed boson and fermion gases. The sign of the thermodynamic curvature in this limit depends on the statistical interaction. In our sign convention, it is positive for boson and $q$-deformed boson gas while it is negative for the others.
\section{Conclusion}
The study of thermodynamic geometry of thermodynamical systems has an archaism as long as three decade. The investigation of known thermodynamic systems has created some application for a new quantity, namely thermodynamic curvature. The thermodynamic curvature is served as a measure of statistical interaction, a sign for phase transition, a tentative measure of stability and in general the critical properties. In this paper, we have endeavored to consider the thermodynamic geometry of symmetric $q$-deformed  bosons and fermions. Therefore, we can argue that the statistical interaction of $q$-deformed boson gas is attractive, while it is repulsive for $q$-deformed fermion gas. Also, due to the stability interpretation of thermodynamic curvature, the $q$-deformed boson gas for large value of deformation parameter ($q<1$) might be more stable than the small value. It means that an ideal boson gas is more stable than the deformed one. Specifically, up to the validity of the stability interpretation, we argue that the $q$-deformed fermion gas could be more stable than the $q$-deformed boson gas for any arbitrary value of deformation parameter. Also, investigation of singular point of thermodynamic curvature represent that the critical fugacity for any arbitrary value of deformation is coincide to the upper bound of fugacity, which is given by the heat capacity definition. Therefore, we can define the phase transition temperature such as Bose-Einstein condensation temperature. We showed that $q$-deformed boson gas ($q\neq 1 $), may condensate in finite temperature in all dimension with particles satisfying the arbitrary dispersion relation. The existence of the finite phase transition temperature for $q$-deformed boson gas even in low dimension is an interesting and new result of present paper.

Obtaining the critical temperature for standard $q$-deformed boson gas \cite{Lavagno2} is straightforward such as previous one and we elude to calculate and represent it because of entire similarity.

In the low temperature limit, we identified a universal behavior for various quantum gases. The thermodynamic curvature is a power-law function with respect to $T$. The exponent is related to the dimension of the system and the dispersion relation of the particles but it is independent of the details of the statistics.

\section*{References}
\bibliographystyle{iopart-num}

\providecommand{\newblock}{}

\end{document}